\documentclass[11pt]{article}

\usepackage[body={17cm, 22.5cm},right=2.2cm]{geometry}

\usepackage{cite}

\usepackage{amssymb,amsmath}

\allowbreak

\newcommand{\be}{\begin{eqnarray}}

\newcommand{\ee}{\end{eqnarray}}

\def\p{\partial}

\def\s{{\sigma}}

\def\nn{\nonumber}

\begin{document}

{\ }
\vspace{-12mm}

\begin{flushright}
	\hfill{DFPD-2017/TH/05} 
\end{flushright}

\baselineskip 5.5 mm

\vspace{0.8mm}

\begin{center}
\vglue 0.10in
{\Large\bf {  Constrained Superfields and Applications }
}

{\ }
\\[0.3cm]
{\large  Fotis Farakos}
\\[0.2cm]

\vspace{.3cm}
{\normalsize {\it  Dipartimento di Fisica e Astronomia ``Galileo Galilei''\\
			Universit\`a di Padova, Via Marzolo 8, 35131 Padova, Italy}}
\\
and
\\
{\normalsize {\it  INFN, Sezione di Padova \\
		Via Marzolo 8, 35131 Padova, Italy}}

\vspace{.3cm}
{\normalsize  E-mail: fotisfgm@gmail.com }

\end{center}

\noindent
{\it Proceedings for Corfu Summer Institute 2016\\  
		``School and Workshops on Elementary Particle Physics and Gravity''\\
                 31 August - 23 September, 2016\\
                 Corfu, Greece  }

\vspace{.4cm}

{
\small  \noindent \textbf{Abstract} \\[0.3cm] 
When supersymmetry is spontaneously broken it will be generically non-linearly realized. 
	A method to describe the non-linear realization of supersymmetry is with constrained superfields. 
	We discuss the basic features of this description and review some recent developments in supergravity.}

\vskip 0.5cm

\def\thefootnote{\arabic{footnote}}
\setcounter{footnote}{0}

\def\ls{\left[}
\def\rs{\right]}
\def\lc{\left\{}
\def\rc{\right\}}

\def\p{\partial}

\def\S{\Sigma}

\def\nn{\nonumber}

\bigskip

\section{Introduction and discussion}

Supersymmetry and supergravity are highly motivated for the study of physics beyond the Standard Model. 
Since we do not observe the supersymmetric partners of the Standard Model particles, 
the breaking of supersymmetry is in the core of any realistic model building scenario. 
The scale of the supersymmetry breaking remains unknown, 
but even if it were known, 
an effective description of the theory in the broken phase would be welcome. 
Non-linear realizations of supersymmetric theories are addressing exactly this: 
how to construct effective theories for systems where supersymmetry is spontaneously broken.

In general when considering a field theory with a spontaneously broken symmetry, 
non-linear realizations can emerge in different manners. 
As a consequence of the symmetry breaking, in the spectrum there are going to be heavy fields characterized by a mass $M$. 
One can decouple then these fields by taking the formal limit of infinite $M$. 
This decoupling is going to implement the non-linear realization on the fields appearing in the low energy theory. 
Alternatively, 
in an energy regime much below the aforementioned mass scale $M$, 
the system can be effectively described with a non-linear realization of the broken symmetry. 
In this case the fields with masses of order $M$ or larger are eliminated from the spectrum.

Superfield methods \cite{Gates:1983nr,Wess:1992cp,Freedman:2012zz}  are especially suited for the study of non-linear realizations  
because they are immediately compatible with the formalism that has been developed 
explicitly for the study of supersymmetry and supergravity. 
An approach that has gained particular attention is the method of {\it constrained superfields}, 
which is the main topic of this article. 
Within this setup there exists a nilpotent chiral superfield $X$, 
which satisfies the constraint \cite{Rocek:1978nb,Lind,Casalbuoni:1988xh} 
\be
\label{X2}
X^2 = 0 \, . 
\ee
This constraint has a non-trivial solution whenever the fermion in $X$ contributes to the goldstino direction. 
In other words, 
when the auxiliary field of $X$, $F^X$, 
sources  the supersymmetry breaking or at least contributes to it.  
When global supersymmetry is spontaneously broken by an F-term or a D-term potential, 
it has been shown that the superfield $X$ always exists \cite{NGF}.

Matter fields in supersymmetric theories reside inside supermultiplets 
and in superspace they are described by superfields. 
When supersymmetry is unbroken, these superfields will contain component fields with degenerate masses. 
Once supersymmetry is broken some component fields will be heavy and will decouple from the system 
if their mass is much larger that the energy scales one is probing. 
The way to describe this decoupling is by imposing constraints of the form \cite{DallAgata:2016syy} 
\be
\label{XXQ}
X \overline X \, Q = 0 \, . 
\ee
These kind of constraints will eliminate the lowest component field of $Q$ from the spectrum  
and it can be shown that this corresponds to the formal decoupling of the eliminated fields \cite{DallAgata:2016syy}. 
By combining various constraints of the form \eqref{XXQ} one can eliminate any heavy component field from the theory. 
This method in fact reproduces all previously known constraints on superfields \cite{Brignole:1997pe,Komargodski:2009rz}. 
These constraints can be generically imposed with the use of Lagrange multipliers \cite{Ferrara:2016een}  
and their study requires the use of Lagrangians before integrating out auxiliary fields, 
which can be found for example in \cite{Gates:1983nr,Wess:1992cp,Freedman:2012zz} 
or \cite{Freedman:2016qnq,Freedman:2017obq}.

Constrained superfields and non-linear realizations have various applications: 
For example the Supersymmetric Standard Model \cite{Antoniadis:2010hs,Farakos:2012fm,Goodsell:2014dia}, 
the current de Sitter phase of our 
universe \cite{Dudas:2015eha,Bergshoeff:2015tra,Hasegawa:2015bza,Ferrara:2015gta,Antoniadis:2015ala,DallAgata:2015pdd,Bandos:2015xnf,Farakos:2016hly,Cribiori:2016qif,Bandos:2016xyu,Benakli:2017whb}, 
or the inflationary phase \cite{Antoniadis:2014oya,Ferrara:2014kva,Kallosh:2014via,DallAgata:2014qsj,Ferrara:2015tyn,Carrasco:2015iij,DallAgata:2015zxp,Kahn:2015mla,Dudas:2016eej,Kallosh:2016ndd,McDonough:2016der,Hasegawa:2017hgd}. 
The String/Brane origin of constrained superfields is also under study \cite{Antoniadis:1999xk,Angelantonj:1999jh,Aldazabal:1999jr,Dudas:2000nv,Pradisi:2001yv,Bergshoeff:2015jxa,Kallosh:2015nia,Garcia-Etxebarria:2015lif,Dasgupta:2016prs,Vercnocke:2016fbt,Kallosh:2016aep,Aalsma:2017ulu}.

In this article we present some technical details of the constrained superfields approach, 
we illustrate the relation to other methods 
and then we turn to inflationary model building in supergravity. 
We use the conventions of  \cite{Wess:1992cp}.

\section{Nonlinear realizations of supersymmetry}

\subsection{Goldstino sector}

When supersymmetry is spontaneously broken, there will always exist a fermionic goldstone field: the goldstino. 
The description of this fermion and its coupling to matter are expected to have universal properties.  
The first effective description for this fermion was provided by Volkov and Akulov in \cite{Volkov:1973ix}.  
In that approach the fermion transforms under supersymmetry as 
\be
\label{VAintro}
\delta \lambda_\alpha =  \xi_\alpha - i \left( \lambda \s^m \overline \xi - \xi \s^m \overline \lambda  \right) \p_m \lambda_\alpha . 
\ee 
The goldstino fermion can be embedded in a spinor superfield $\Lambda_\alpha$ ($\lambda_\alpha=\Lambda_\alpha|$), 
which can be used to describe 
its couplings to matter superfields \cite{IK1,Samuel:1982uh}. 
The superfield $\Lambda_\alpha$ satisfies the constraints 
\be
\label{superVA}
\begin{aligned}
D_\beta \Lambda_\alpha  &= \epsilon_{\alpha \beta} + i \sigma_{\beta \dot \beta}^{m} \overline \Lambda^{\dot \beta} \p_m \Lambda_\alpha \, , 
\\
\overline D_{\dot \beta} \Lambda_\alpha &= -  i \, \Lambda^\beta \sigma_{\beta \dot \beta}^{m}  \p_m \Lambda_\alpha \, , 
\end{aligned}
\ee
and the only independent component inside it is the fermion $\lambda_\alpha$. 
A simple redefinition, 
relates the superfield $\Lambda_\alpha$ to an alternative version, 
which is the superfield $\Gamma_\alpha$, and this one satisfies the constraints 
\be
\begin{aligned}
D_\alpha \Gamma_\beta  &= \epsilon_{\beta \alpha} \, , 
\\
\overline D^{\dot \beta} \Gamma^\alpha &= 2 i \, \left( \overline \s^m \, \Gamma \right)^{\dot \beta} \, \p_m \Gamma^\alpha \, .  
\end{aligned}
\ee
The only independent component field in $\Gamma_\alpha$ is the fermion $\gamma_\alpha$ ($\gamma_\alpha=\Gamma_\alpha|$), 
which transforms under supersymmetry as 
\be
\label{gammaintro}
\delta \gamma_\alpha =  \xi_\alpha 
+ 2 i \, \xi \s^m \overline \gamma \,  \p_m \gamma_\alpha  . 
\ee 
The redefinitions relating $\lambda_\alpha$ to $\gamma_\alpha$ can be already found in \cite{Samuel:1982uh}. 
The superspace relation is 
\be
\Gamma_\alpha = -2 \frac{D_\alpha \overline D^2 
\left( \Lambda^2 \overline \Lambda^2 \right)}{D^2 \overline D^2 
\left(\Lambda^2 \overline \Lambda^2\right)} \, . 
\ee

As we noted in the introduction, 
an alternative description for the embedding of the goldstino into a superfield is provided by a 
constrained chiral superfield $X$ which satisfies the constraint \eqref{X2}. 
The superspace expansion of $X$ will then be \cite{Casalbuoni:1988xh} 
\be
X = \frac{G^2}{2 F^X} + \sqrt 2 \theta G + \theta^2 F^X \, . 
\ee
The goldstino resides in the fermion component field of $X$, $G_\alpha$, 
and this formulation is consistent only if the vacuum expectation value of the  auxiliary field of the $X$ superfield is non-vanishing: 
$\langle F^X \rangle \ne 0$. 
The $\Gamma_\alpha$ and $X$ formalisms can be related via \cite{Cribiori:2016hdz} 
\be
\label{GXGX}
\Gamma_\alpha = -2 \frac{D_\alpha X}{D^2 X} . 
\ee
The equivalence of these formulations for the free theories 
was shown in component form in \cite{Kuzenko:2011tj} and later in full superspace in \cite{Cribiori:2016hdz}. 
In \cite{NGF} it has been shown that the equivalence holds also for matter couplings. 
Therefore, 
any Lagrangian which would contain $\gamma_\alpha$ can be written as a Lagrangian of $X$, 
by replacing in the matter couplings 
\be
\gamma_\alpha = \frac{G_\alpha}{\sqrt 2 F^X}  
\ee
and introducing the appropriate $X$ sector as has been shown in \cite{NGF}.

Supersymmetric Lagrangians with chiral superfields are usually constructed in superspace as 
\be
\label{susyL}
{\cal L} = \int d^4 \theta \, K + \left( f \int d^2 \theta \, W + c.c.  \right) \, , 
\ee
where $K$ is the K\"ahler potential, which is a hermitian function of the chiral superfields, 
and $W$ the superpotential, which is a holomorphic function of the chiral superfields. 
The simplest Lagrangian we can construct with the constrained superfield $X$ has 
\be
K = X \overline X \ , \ W = f \, X \, . 
\ee
This Lagrangian in component form gives (after we integrate out $F^X$) 
\be
\label{XXVA}
{\cal L} = 
- f^2 + i \p_m \overline G \overline \s^m G + \frac{1}{4 f^2} \overline G^2 \p^2 G^2 
- \frac{1}{16 f^6} G^2 \overline G^2 \p^2 G^2 \p^2 \overline G^2 \, . 
\ee
The supersymmetry transformation of the goldstino will read 
\be
\delta G_\alpha = - f \xi_\alpha - (i/2 f)  \sigma^{m}_{\alpha \dot \alpha} \overline \xi^{\dot \alpha} \partial_m G^2 + \cdots
\ee
The equivalence of the Lagrangian \eqref{XXVA} to the Volkov--Akulov model can been proved in various ways \cite{Kuzenko:2011tj,Cribiori:2016hdz}. 
Notice that this theory contains higher derivatives 
due to the non-linear realization of supersymmetry and that the number of fermions and bosons is manifestly not the same.

In \cite{NGF} it has been shown that when supersymmetry is spontaneously broken either from an F-term 
or a D-term (or both), then the low energy theory will be always described by the nilpotent chiral superfield $X$ 
and the leading terms of the goldstino sector will have the form \eqref{XXVA}.

Supersymmetry can be broken also from complex linear multiplets 
\cite{Kuzenko:2011ti,Farakos:2013zsa,Farakos:2015vba,Koci:2016rqf}. 
These theories generically reproduce the Lagrangian \eqref{XXVA} for the goldstino sector.

\subsection{Matter sector}

To describe the matter sector we use superfields. 
When we discuss an effective theory where supersymmetry is spontaneously broken, 
some component fields will have masses much larger than the energy scales we are probing 
and therefore may be eliminated from the spectrum. 
The way to eliminate them is by imposing constraints of the form \eqref{XXQ} 
on the matter superfields. 
A complete discussion of the properties of these constraints can be found in \cite{DallAgata:2016syy}.  
Here we will just treat a few simple examples. 

Assume we have a low energy theory where, 
on top of the superfield $X$, 
we have  the chiral superfield $Y$ with superspace expansion 
\be 
Y = y + \sqrt 2 \theta \chi^Y + \theta^2 F^Y \, . 
\ee 
If the scalar in $Y$ is heavy we can eliminate it from the spectrum and, 
by using the prescription of \cite{DallAgata:2016syy}, 
we can impose the constraint 
\be
\label{XXY}
X \overline X \, Y = 0 \, . 
\ee 
The solution to the superspace constraint \eqref{XXY} will then simply be 
\be
y = \frac{G \chi^Y}{F^X} - \frac{G^2}{2 F^2} F^Y \, . 
\ee
This constrained superfield was originally studied in the simpler (but equivalent) form $XY=0$ in \cite{Brignole:1997pe,Komargodski:2009rz}.

One can also construct matter supermultiplets which contain only a single independent component field \cite{Komargodski:2009rz}. 
Consider a chiral superfield ${\cal A}$, 
with superspace expansion 
\be
{\cal A} = \varphi + i  b + \sqrt 2 \theta \chi^A + \theta^2 F^A \, , 
\ee
and let us show how to keep only the real scalar $\varphi$ in the theory. 
To this end we impose a series of constraints 
\be
\label{XAXA00}
\begin{aligned}
& |X|^2 \left( {\cal A} - \overline {\cal A} \right) = 0 \, , 
\\
& |X|^2 \, D_\alpha {\cal A} = 0 \, , 
\\
& |X|^2 \, D^2 {\cal A} = 0 \, . 
\end{aligned}
\ee
Once we solve these constraints all the component fields of ${\cal A}$ are eliminated and only $\varphi$ remains as independent. 
This is evident from the fact that the constraint $X \overline X Q=0$ eliminates the component field $Q|$. 
The complete solution is 
\be
\begin{aligned}
\chi^A & = i \sigma^m  \left( \frac{\overline G }{\overline F}  \right) \partial_m (\varphi + i b)  \, , 
\\
F^A & = \left( \frac{\overline G^2}{2 \overline F^2} \partial^2 (\varphi + i b)  
- \partial_n \left( \frac{\overline G}{\overline F}  \right) \overline \sigma^m \sigma^n \frac{\overline G}{\overline F}  \partial_m (\varphi + i b)    \right) \, , 
\end{aligned}
\ee
where 
\be
\begin{aligned}
b = & \frac12 \left( \frac{G}{F} \s^m \frac{\overline G}{\overline F} \right) \p_m \varphi  
- \left( \frac{i}{8} \frac{G^2}{F^2} \p_n \left( \frac{\overline G}{\overline F} \right) 
\overline \s^m \s^n 
\frac{\overline G}{\overline F} \p_m \varphi \, + c.c. 
\right) 
\\
& 
- \frac{G^2 \overline G^2}{32 \, F \overline F} 
\p_n \left( \frac{\overline G}{\overline F} \right)  
(\overline \s^p \s^n \overline \s^m  + \overline \s^n \s^m \overline \s^p  ) 
\p_m \left( \frac{G}{F} \right) \p_p \varphi \, . 
\end{aligned}
\ee
This constrained superfield has been first studied in the form $X {\cal A} = X \overline {\cal A}$ in \cite{Komargodski:2009rz}. 
It has been shown in \cite{DallAgata:2016syy} that imposing the constraints \eqref{XAXA00} is equivalent to imposing $X {\cal A} = X \overline {\cal A}$.

We should warn the reader that when imposing constraints of the form \eqref{XXQ} 
inconsistencies might arise if the component field to be eliminated satisfies some sort of Bianchi identities. 
For example the constraint \eqref{XXQ}  cannot be used to eliminate directly a field strength $F_{mn}$, 
because the solution might violate the algebraic condition $\p_{[k}F_{mn]}=0$. 
The consistent way to eliminate gauge fields has been outlined in \cite{DallAgata:2016syy}. 
Essentially one has to include the degrees of freedom that make the gauge field massive and then decouple it. 
This is in accordance with the fact that these constraints are equivalent to the decoupling of the heavy massive fields.

It is instructive at this point  to study the relation between the various formulations for describing matter. 
When the goldstino resides in the superfield $\Lambda_\alpha$, which is defined by \eqref{superVA}, 
it is known how multiplets with a single independent component field can be built \cite{Wess:1992cp}. 
The method is the following: Start from an unconstrained superfield ${\cal P}$, 
and impose the constraints 
\be
\label{WBC}
\begin{aligned}
D_\alpha {\cal P} & = i \s^m_{\alpha \dot \alpha} \overline \Lambda^{\dot \alpha} \p_m {\cal P} \, , 
\\
\overline D_{\dot \alpha} {\cal P} & = - i \Lambda^\alpha \s^m_{\alpha \dot \alpha} \p_m {\cal P} \, . 
\end{aligned}
\ee
Then the only independent component in ${\cal P}$ will be $p= {\cal P}|$. 
The other component fields in ${\cal P}$ will be functions of $\lambda_\alpha$ and $p$. 
Now notice that the constraints \eqref{WBC} also imply 
\be
\Lambda^2 \overline \Lambda^2 \, D_\beta {\cal P} = 0 \, , 
\quad 
\Lambda^2 \overline \Lambda^2 \, \overline D_{\dot \beta} {\cal P} = 0 \, , 
\quad 
\Lambda^2 \overline \Lambda^2 \, D^2 {\cal P} = 0 \, , 
\quad \cdots  
\ee
where the dots stand for the rest of the constraints arising from  \eqref{WBC} 
which are all of the form 
\be 
\label{LLQ}
\Lambda^2 \overline \Lambda^2 Q=0 \, . 
\ee
Relating now $\Lambda_\alpha$ to $X$ via \eqref{GXGX} we have  
\be 
\Lambda^2 \overline \Lambda^2 \sim \Gamma^2 \overline \Gamma^2 \sim X \overline X \, , 
\ee
which eventually links \eqref{LLQ} to \eqref{XXQ}. 
In this way we see exactly how the two formalisms are related.

\section{Nonlinear realizations of local supersymmetry}

In this section we discuss non-linear realizations of supersymmetry in supergravity. 
We will focus on the method of constrained superfields, 
but will comment on the relation to other formalisms when it is instructive. 
Finally we will discuss applications to inflation.

We will not review supergravity here, 
but we can remind the reader that 
the generic coupling of chiral superfields to the old-minimal supergravity is constructed with the superspace Lagrangian 
\begin{equation}
\label{L1}
{\cal L} = \int d^2 \Theta \, 2 {\cal E}\, \left[ \frac38 \left( \overline {\cal D}^2 - 8 {\cal R} \right) \, \text{e}^{-K/3} + W \right] + c.c. \, , 
\end{equation} 
where we set $M_P=1$, 
and the chiral density $2 {\cal E}$ has superspace expansion 
\be
2 {\cal E} = e \left\{ 1+ i\Theta \sigma^{a} \overline{\psi}_a 
-\Theta^2 \Big{(} \overline M +\overline{\psi}_a \overline{\sigma}^{ab}\overline{\psi}_b 
\Big{)} \right\} \, . 
\ee
The Ricci superfield ${\cal R}$ is a chiral superfield $\overline{ {\cal D}}_{\dot{\alpha}} {\cal R} =0$, 
whose lowest component is the auxiliary field $M$ of the old-minimal supergravity multiplet 
\be
{\cal R} | = -\frac{1}{6} M \, .
\ee
The fermionic component of ${\cal R}$  is 
\be
{\cal D}_{\alpha} {\cal R} | = 
-\frac{1}{6} ( \s^a \overline \s^b \psi_{a b} + i b^a \psi_{a } 
-i \s^a \overline \psi_{a } M )_{\alpha} \, , 
\ee
where $\psi_m^\alpha$ is the gravitino, the superpartner of the gravitational field $e_m^a$, and $\psi_{mn}^\alpha$ is its field strength. 
The  highest component of ${\cal R}$ is 
\be
\begin{split}
{\cal D}^2 {\cal R} | =&  -\frac{1}{3} R + \frac{4}{9} M \overline M + \frac{2}{9} b^a b_a 
-\frac{2 i}{3} e_{a}^{\ m} {\cal D}_m b^a 
+\frac{1}{3} \overline \psi \overline \psi M -\frac{1}{3} \psi_m \s^m \overline \psi_n b^n 
\\
&+ \frac{2 i}{3} \overline \psi^m \overline \s^n \psi_{mn} 
+ \frac{1}{12} \epsilon^{k l m n} [ \overline \psi_k \overline \s_l \psi_{mn} + \psi_k \s_l \overline \psi_{mn}]  \, .
\end{split}
\ee
The real vector $b_a$ is an auxiliary field of the old-minimal supergravity multiplet.

\subsection{Constrained superfields in supergravity}

If superymmerty is broken by a chiral superfield $X$, 
then even in supergravity we can impose the constraint \eqref{X2} 
and the superfield $X$ becomes 
\be
X = \frac{G^2}{2 F^X} + \sqrt 2 \Theta G + \Theta^2 F^X \, . 
\ee
Moreover, because the goldstino {\it is} a goldstone mode, 
when supersymmetry is broken it will be absorbed by the gravitino which will become massive. 
This means that since the goldstino is a pure gauge degree of freedom, 
and since the $G_\alpha$ fermion by assumption always contributes to the goldstino, we can always fix it to 
\be
\label{GG00} 
G_\alpha =0 \, , 
\ee
in the final Lagrangian. 
We will use the gauge \eqref{GG00} for the rest of the article when we write down Lagrangians in component form. 
Note that the gauge choice \eqref{GG00} might not be always the unitary gauge.

The simplest model which breaks supersymmetry in supergravity has a flat K\"ahler potential 
\be
\label{KXX}
K = X \overline X \, , 
\ee
and a superpotential 
\be
\label{WXX}
W = f \, X + W_0 \, . 
\ee
Here $f$ and $W_0$ are complex constants. 
Once we insert \eqref{KXX} and \eqref{WXX} into \eqref{L1}, 
the Lagrangian of supergravity coupled to $X$ takes the component form \cite{Dudas:2015eha,Bergshoeff:2015tra,Hasegawa:2015bza} 
\begin{equation} 
\label{freeDS}
\begin{split}
e^{-1} {\cal L} = &  -\frac12 R 
+ \frac{1}{2} \epsilon^{klmn} (\overline \psi_k \overline \sigma_l {\cal D}_m \psi_n - \psi_k \sigma_l {\cal D}_m \overline \psi_n) 
\\[2mm]
&  
- W_0 \, \overline \psi_a \overline \sigma^{ab} \overline \psi_b 
- \overline{W_0} \, \psi_a  \sigma^{ab}  \psi_b 
- \Lambda  \, . 
\end{split}
\end{equation}
In \eqref{freeDS} supersymmetry is spontaneously broken on a vacuum that can be Minkowski, de Sitter or anti-de Sitter according to the value of the $f$ and $W_0$ constants, 
which govern the cosmological constant 
\be
\Lambda = |f|^2 - 3 |W_0|^2 \, . 
\ee

The description of matter constrained superfields works in the same way as in supersymmetry. 
One can use the generic constraint 
\be
X \overline X Q =0 \, , 
\ee
to eliminate the heavy component fields from any given superfield. 
The proof that this procedure is equivalent to the standard component form procedure for local supersymmetry 
can be found in \cite{Cribiori:2016qif}.

We can also follow the approach where the goldstino resides into the spinor superfield $\Gamma_\alpha$. 
In local supersymmetry we have 
\be
\begin{aligned}
{\cal D}_\alpha \Gamma_\beta  &= \epsilon_{\beta \alpha} \left( 1 - 2 \, \Gamma^2 {\cal R} \right)  \, , 
\\
\overline{\cal D}^{\dot \beta} \Gamma^\alpha &= 2 i \, \left( \overline \s^a \, \Gamma \right)^{\dot \beta} \, {\cal D}_a \Gamma^\alpha 
+ \frac12 \, \Gamma^2 B^{\dot \beta \alpha} \, .  
\end{aligned}
\ee
Here $B_a$ is a superfield which has lowest component $B_a| = -b_a/3$. 
The relation between $\Gamma_\alpha$ and $X$ is given again by \eqref{GXGX}, 
but now the superspace derivatives become covariant  \cite{Bandos:2016xyu} 
\be
\Gamma_\alpha = -2 \frac{{\cal D}_\alpha X}{{\cal D}^2 X} \, . 
\ee
In terms of $\Gamma_\alpha$ the pure goldstino 
sector is\footnote{We use $\int d^4 \theta \, E \, U= -\frac18 
\int d^2 \Theta \, 2 {\cal E}\, \left[ \left( \overline {\cal D}^2 - 8 {\cal R} \right) U \right] +c.c.$ 
for a hermitian superfield $U$.} 
\be 
\label{GGGGGG}
{\cal L} =  - \frac{1}{16 \, \kappa^2} \int d^4 \theta \, E \, \Gamma^2 \overline \Gamma^2 \, ,  
\ee
where $\kappa^{-1/2}$ is a scale which will enter the goldstino interactions once it is coupled to matter.

We would like to bring to the reader's attention that 
a local formulation of the geometric Volkov--Akulov approach is provided by the {\it goldstino brane} \cite{Bandos:2015xnf,Bandos:2016xyu}. 
The free goldstino sector in this setup, when coupled to supergravity, reproduces \eqref{freeDS}.

\subsection{Constrained supergravity}

Constraints can be also imposed on the auxiliary fields of the gravity multiplet. 
For example one can simply impose a constraint of the form \cite{Cribiori:2016qif} 
\be
\label{wawawewa}
\overline X X \left( {\cal R} + \frac{c}{6}  \right) = 0 \, , 
\ee
where $c$ is a complex constant. 
We will refer here to the supergravity theory satisfying \eqref{wawawewa} as {\it constrained supergravity}. 
Solving the constraint \eqref{wawawewa} delivers  
\be
M = c + {\cal O}(G, \overline G) \, . 
\ee
To build a minimal theory within this setup, 
consider the Lagrangian 
\begin{equation} 
{\cal L} =  \int d^2 \Theta \, 2 {\cal E}\, \left[ \frac38 \left( \overline {\cal D}^2 - 8 {\cal R} \right) \, \text{e}^{-|X|^2/3} + (f \, X + W_0) \right] + c.c. \, , 
\end{equation} 
and by imposing the constraint \eqref{wawawewa} 
one finds the Lagrangian \eqref{freeDS} with a cosmological constant given by 
\begin{equation}
	\Lambda = \frac13 |c|^2+ |f|^2 + m_{3/2} \overline c +\overline m_{3/2} c = \Lambda_S - 3 |m_{3/2}|^2\, . 
\end{equation} 
Notice that here the supersymmetry breaking contribution to the vacuum energy is 
\begin{equation}
	\label{LambdaS}
	\Lambda_S = |f|^2 + \left|\frac{c}{\sqrt3} + \sqrt3\, m_{3/2}\right|^2 \, .
\end{equation} 
The interested reader can find more details in \cite{Cribiori:2016qif}, where the properties of these theories are studied in superspace. 
Similar  models can be found in \cite{Delacretaz:2016nhw} constructed with the standard component form procedure.

We would like here to study the relation of these theories to standard supergravity. 
To this end we impose the constraint \eqref{wawawewa} with a chiral superfield Lagrange multiplier ${\cal M}$ and then study the dual theory. 
To impose the constraint we introduce the term 
\be
\label{LLMM}
{\cal L}_c = \int d^2 \Theta \, 2 {\cal E} \, {\cal M} X   \left( {\cal R} + \frac{c}{6}  \right)  + c.c. 
\ee
If we vary the chiral superfield $\cal M$ we get the constraint on the supergravity multiplet \eqref{wawawewa}. 
To find the dual theory we first rewrite \eqref{LLMM} as 
\be
\label{newLLMM}
{\cal L}_c = -\frac18 \int d^2 \Theta \, 2 {\cal E} \left( \overline{\cal D}^2 - 8 {\cal R} \right)    
\Big{[} {\cal M} X +  \overline{\cal M} \overline X \Big{]} 
+ \frac16 \left( \int d^2 \Theta \, 2 {\cal E} \, c \,  {\cal M} X + c.c. \right) \, . 
\ee
If we now add \eqref{newLLMM} to the standard goldstino sector, 
we have in total 
\begin{equation}
\label{LMXMX}
\begin{aligned}
{\cal L} = & \int d^2 \Theta \, 2 {\cal E}\, \left[ \frac38 \left( \overline {\cal D}^2 - 8 {\cal R} \right) \, \text{e}^{-|X|^2/3} + (f \, X + W_0) \right] + c.c. \, 
\\
& -\frac18 \int d^2 \Theta \, 2 {\cal E} \left( \overline{\cal D}^2 - 8 {\cal R} \right)    
\Big{[} {\cal M} X +  \overline{\cal M} \overline X \Big{]} 
+ \frac16 \left( \int d^2 \Theta \, 2 {\cal E} \, c \,  {\cal M} X + c.c. \right) \, . 
\end{aligned}
\end{equation} 
Notice that we have found standard supergravity coupled to 
\be
\begin{aligned}
K &= - 3 \, \text{ln} \left( 1 -  \frac{X \overline X}{3} - \frac{{\cal M} X}{6} - \frac{\overline {\cal M} \overline X}{6} \right) \, , 
\\
W &= W_0 + f X + \frac{c}{6} {\cal M} X \, ,  
\end{aligned}
\ee
which after a  K\"ahler transformation can be rewritten as standard supergravity coupled to 
\be
\label{VAR}
\begin{aligned}
K &= - 3 \, \text{ln} \left( 1 -  \frac{X \overline X}{3} \Big{[} 1 + \frac{{\cal M} \overline {\cal M}}{12} \Big{]}  \right) \, , 
\\
W &= W_0 + f X + \left( \frac{W_0}{2} +  \frac{c}{6} \right) {\cal M} X \, . 
\end{aligned}
\ee
Now we vary the supergravity theory coupled to \eqref{VAR} with respect to ${\cal M}$, 
and we find 
\be
\label{Monsh1}
- \frac14 \left( \overline{\cal D}^2 - 8 {\cal R} \right)  X \overline X \, \frac{\overline{\cal M}}{12} = 
- X \left( \frac{W_0}{2} +  \frac{c}{6} \right) , 
\ee
which also gives 
\be
\label{Monsh2}
\frac{1}{12} X \overline X \, {\cal M} \overline {\cal M} = 
 12 |X|^2  \frac{16}{|{\cal D}^2 X |^2} \Big{|} \frac{W_0}{2} +  \frac{c}{6}  \Big{|}^2 . 
\ee 
Since $\cal M$ represents only {\it auxiliary} degrees of freedom it can be consistently integrated out from the Lagrangian 
- it is after all a Lagrange multiplier. 
Indeed, 
using equations \eqref{Monsh1} and \eqref{Monsh2} we find that the total Lagrangian becomes 
\be
\label{HDHD}
{\cal L} = \int d^4 \theta E \left( -3 + X \overline X - 12 |X|^2  \frac{16}{|{\cal D}^2 X |^2} \Big{|} \frac{W_0}{2} +  \frac{c}{6}  \Big{|}^2 \right) 
+ \left( \int d^2 \Theta \, 2 {\cal E} \left( W_0 + f X \right) + c.c. \right) . 
\ee

We thus find that 
the minimal Lagrangian of constrained supergravity coupled to $X$ 
is {\it on-shell} 
equivalent to the Lagrangian \eqref{HDHD}, 
which does not have the form of standard supergravity since it contains superspace higher derivatives. 
We stress here that we only proved an on-shell equivalence for the free theories, 
therefore other properties, 
as for example matter couplings, 
will in principle be different.

The superspace higher derivative term in the Lagrangian \eqref{HDHD} contributes to the vacuum energy and to the goldstino kinetic term. 
To see how this happens, we will bring the Lagrangian \eqref{HDHD} to an 
(on-shell) equivalent form where there are no higher derivative terms. 
In this way we can also uncover exactly how the unitarity bound on the gravitino mass is related to the parameters of the theory. 
To do so we have to integrate out the auxiliary field $F^X$, 
so that the two terms containing $|X|^2$ in \eqref{HDHD} get the same form. 
This can be done by varying $X$ but multiplying the equation with $|X|^2$. 
This procedure gives 
\be
\label{LR}
\frac{16 \, |X|^2}{|{\cal D}^2 X|^2} = \frac{|X|^2}{|f|^2}. 
\ee
Now the goldstino superfield $X$ satisfies both constraints 
\eqref{X2} and \eqref{LR}. 
The $X^2=0$ constraint eliminates the lowest scalar component of $X$ 
and the new constraint \eqref{LR} eliminates the auxiliary field of $X$, $F^X$. 
These constraints were originally studied in \cite{Lind}. 
Due to these constraints on $X$ the Lagrangian takes a simpler form 
\be
{\cal L} = \int d^4 \theta E \left( -3 - X \overline X \Big{[} 1 + \Big{|} \sqrt 3 \, W_0 +  \frac{c}{\sqrt 3}  \Big{|}^2 / \, | f |^2 \Big{]}  \right) 
+ \left( \int d^2 \Theta \, 2 {\cal E} \, W_0  + c.c. \right) . 
\ee
We now identify $\Lambda_S$ from \eqref{LambdaS}, 
and the Lagrangian becomes 
\be
{\cal L} = \int d^4 \theta E \left( -3 - \frac{\Lambda_S}{|f|^2} \, X \overline X \right) 
+ \left( \int d^2 \Theta \, 2 {\cal E} \, W_0  + c.c. \right) . 
\ee 
Finally, since we are interested in an on-shell equivalence we can relax the constraint \eqref{LR} on the auxiliary field $F^X$, 
and the Lagrangian takes the form of  
standard supergravity coupled to the nilpotent $X$ superfield 
\be
{\cal L} = \int d^4 \theta E \left( -3 + \frac{|f|^2}{\Lambda_S} X \overline X  \right) 
+ \left( \int d^2 \Theta \, 2 {\cal E} \left( W_0 + f X \right) + c.c. \right)  \ , \quad X^2=0 \, . 
\ee
The goldstino kinetic term has the canonical sign when 
\be
\label{bound}
\Lambda_S > 0 \, . 
\ee 
The parameter region \eqref{bound} is exactly the one where the massive gravitino is unitary. 
Indeed, the vacuum energy here is 
\be
\Lambda = \Lambda_S - 3 |W_0 |^2  \, ,  
\ee
and the gravitino mass is 
\be
m_{3/2} = W_0 \, . 
\ee
Notice that the bound \eqref{bound} is automatically satisfied in the models of \cite{Cribiori:2016qif} due to \eqref{LambdaS}.

Let us close this part with a comment on the gravitino mass. 
We see that in both the models of \cite{Cribiori:2016qif} and of \cite{Delacretaz:2016nhw}, 
the relation of the gravitino mass to the vacuum energy is not as transparent as in standard supergravity. 
We would like to stress that the first time models with this property were constructed in superspace 
was in \cite{Bandos:2016xyu}, 
in the context of the goldstino brane, 
where an independent gravitino mass term was introduced.

\subsection{Effective supergravity models for inflation}

The inflationary paradigm postulates that the early universe underwent a period of accelerated expansion. 
The simplest realization of the inflationary paradigm in a field theory setup is {\it single field} inflation, 
where the potential energy of $\varphi$ (the {\it inflaton}) dominates the energy density of our universe. 
The coupling of the inflaton to gravity is 
\be
e^{-1} {\cal L} = - \frac12 M_P^2 \, R - \frac12 \p \varphi \p \varphi - {\cal V}(\varphi) \, , 
\ee 
and inflation takes place when 
\be
\nn
\epsilon = \frac12 M_P^2 \left( \frac{{\cal V}'}{{\cal V}} \right)^2  \ll1 
\, , \quad 
| \eta | = M_P^2 \Big{|} \frac{{\cal V}''}{{\cal V}} \Big{|} \ll 1 . 
\ee 
The parameters $\epsilon$ and $\eta$ are referred to as {\it slow-roll} parameters. 
The current constraints on single field inflation from the Planck collaboration give roughly 
\be
\nn
6 \epsilon - 2 \eta \sim 0.032 
\, , \quad 
16 \, \epsilon < 0.12 \, , 
\ee
where the slow-roll parameters are to be evaluated at $50$ to $60$ e-foldings from the end of inflation. 
The scale of inflation is roughly at the GUT scale or a few orders of magnitude lower. 

If supergravity is relevant at the energy scales in which inflation takes place, 
then it has to be used for the description of the inflationary phase of our universe. 
We will  give a simple example of how the embedding of inflation in supergravity can be achieved with the use of constrained superfields. 
The interested reader can find a recent review of the topic in \cite{Ferrara:2016ajl}.

Following \cite{Kahn:2015mla,Ferrara:2015tyn,Carrasco:2015iij,DallAgata:2015zxp} we introduce a chiral superfield 
\begin{equation}
{\cal A} = \varphi + i \, b + \sqrt 2 \Theta^\alpha \psi^A_\alpha + \Theta^2 F^A \, , 
\end{equation}
and couple it to the nilpotent goldstino superfield $X$, requiring that \cite{Komargodski:2009rz} 
\begin{equation}
\label{XA}
X {\cal A} - X \overline {\cal A} = 0 . 
\end{equation} 
The most general coupling of ${\cal A}$ with the nilpotent 
superfield $X$ in supergravity has been studied in \cite{Ferrara:2015tyn,Carrasco:2015iij,DallAgata:2015zxp}. 
Supersymmetry is essentially broken by $\langle F^X \rangle \neq 0$ and we can use the $G_{\alpha} =0$ gauge, such that 
\begin{equation}
X|_{G=0} = \,  \Theta^2 F^X \, , \qquad {\cal A}|_{G=0} = \, \varphi \, .  
\end{equation} 
To give a simple example we choose a K\"ahler potential and superpotential of the form 
\begin{equation}
\begin{split}
K =& X \overline X - \frac14 ({\cal A} - \overline {\cal A})^2 \, ,
\\
W =& g({\cal A}) + X f({\cal A}) \, ,
\end{split}
\end{equation}
where $\overline{ f(z)} = f (\overline z)$ and $\overline{ g(z)} = g (\overline z)$. 
The complete theory in the $G_{\alpha}=0$ gauge is described by \cite{Kahn:2015mla,Ferrara:2015tyn,Carrasco:2015iij,DallAgata:2015zxp} 
\begin{equation} 
\label{LLAA}
\begin{split}
e^{-1} {\cal L} = & -\frac12 R
+ \frac{1}{2} \epsilon^{klmn} (\overline \psi_k \overline \sigma_l {\cal D}_m \psi_n - \psi_k \sigma_l {\cal D}_m \psi_n)
\\[2mm]
&
- \frac12 \partial^m \varphi \, \partial_m \varphi  
- g(\varphi) ( \overline \psi_a \overline \sigma^{ab} \overline \psi_b + \psi_a \sigma^{ab} \psi_b ) 
- {\cal V}(\varphi) \, , 
\end{split}
\end{equation}
where the scalar potential is 
\begin{equation}
{\cal V}(\varphi) = f^2(\varphi) - 3 \, g^2(\varphi).
\end{equation}
The scalar potential can be arbitrarily fixed in terms of the two functions $f$ and $g$, 
to drive inflation and to give the desired vacuum energy at the end of inflation, 
though one should carefully choose them in order for the effective theory 
to remain valid in the large range of scales touched during inflation and its exit period. 
Finally we stress that this description is only valid when 
\be
\langle F^X \rangle \ne 0 \, , 
\ee
which here translates to 
\be
\langle f(\varphi) \rangle \ne 0 \, . 
\ee

Let us close this part by reporting a very recent development in the field. 
It has been shown in \cite{Hasegawa:2017hgd} that, unless the function $g$ entering the superpotential 
is very tuned, the models \eqref{LLAA} might suffer from an explosive gravitino production after inflation.

\bigskip

\section*{Acknowledgments}

\noindent 
N. Cribiori, G. Dall'Agata, E. Dudas and M. Porrati are gratefully acknowledged for discussions and collaborations. 
This work is supported in parts by the Padova University Project CPDA119349.


{\small 




\begin{thebibliography}{99}



 
\bibitem{Gates:1983nr} 
  S.~J.~Gates, M.~T.~Grisaru, M.~Rocek and W.~Siegel,
  ``Superspace Or One Thousand and One Lessons in Supersymmetry,''
  Front.\ Phys.\  {\bf 58}, 1 (1983)
  [hep-th/0108200].


\bibitem{Wess:1992cp} 
  J.~Wess and J.~Bagger,
  ``Supersymmetry and supergravity,''
  Princeton, USA: Univ. Pr. (1992). 




\bibitem{Freedman:2012zz} 
  D.~Z.~Freedman and A.~Van Proeyen,
  ``Supergravity,'' 
  Cambridge University Press (2012). 
 
  


 
 
 



\bibitem{Rocek:1978nb} 
  M.~Rocek,
  Phys.\ Rev.\ Lett.\  {\bf 41}, 451 (1978).





\bibitem{Lind}
U. Lindstrom and M. Rocek, 
Phys. Rev. D \textbf{19} (1979) 2300.





 
\bibitem{Casalbuoni:1988xh} 
  R.~Casalbuoni, S.~De Curtis, D.~Dominici, F.~Feruglio and R.~Gatto,
  Phys.\ Lett.\ B {\bf 220}, 569 (1989).



 
 
\bibitem{NGF} 
  N.~Cribiori, G.~Dall'Agata and F.~Farakos,
  arXiv:1704.07387 [hep-th].



 
\bibitem{DallAgata:2016syy}
  G.~Dall'Agata, E.~Dudas and F.~Farakos,
  JHEP {\bf 1605} (2016) 041
  [arXiv:1603.03416 [hep-th]].
 
 
 
 
 
 

\bibitem{Brignole:1997pe} 
  A.~Brignole, F.~Feruglio and F.~Zwirner,
  JHEP {\bf 9711}, 001 (1997)
  [hep-th/9709111].


\bibitem{Komargodski:2009rz} 
  Z.~Komargodski and N.~Seiberg,
  JHEP {\bf 0909}, 066 (2009)
  [arXiv:0907.2441 [hep-th]].


 
 
 
\bibitem{Ferrara:2016een} 
  S.~Ferrara, R.~Kallosh, A.~Van Proeyen and T.~Wrase,
  JHEP {\bf 1604}, 065 (2016)
  [arXiv:1603.02653 [hep-th]].







 
\bibitem{Freedman:2016qnq} 
  D.~Z.~Freedman, D.~Roest and A.~Van Proeyen,
  arXiv:1609.07362 [hep-th].

 
\bibitem{Freedman:2017obq} 
  D.~Z.~Freedman, D.~Roest and A.~Van Proeyen,
  JHEP {\bf 1702}, 102 (2017)
  [arXiv:1701.05216 [hep-th]].



 
 
 
 
 
\bibitem{Antoniadis:2010hs} 
  I.~Antoniadis, E.~Dudas, D.~M.~Ghilencea and P.~Tziveloglou,
  Nucl.\ Phys.\ B {\bf 841}, 157 (2010)
  [arXiv:1006.1662 [hep-ph]].


\bibitem{Farakos:2012fm} 
  F.~Farakos and A.~Kehagias,
  Phys.\ Lett.\ B {\bf 719}, 95 (2013)
  [arXiv:1210.4941 [hep-ph]].
 
 
 
\bibitem{Goodsell:2014dia} 
  M.~D.~Goodsell and P.~Tziveloglou,
  Nucl.\ Phys.\ B {\bf 889}, 650 (2014)
  [arXiv:1407.5076 [hep-ph]].
 
 




\bibitem{Dudas:2015eha}    
E.~Dudas, S.~Ferrara, A.~Kehagias and A.~Sagnotti,    
JHEP {\bf 1509} (2015) 217    
[arXiv:1507.07842 [hep-th]].

\bibitem{Bergshoeff:2015tra}  
E.~A.~Bergshoeff, D.~Z.~Freedman, R.~Kallosh and A.~Van Proeyen, 
Phys.\ Rev.\ D {\bf 92} (2015) no.8,  085040   Erratum: [Phys.\ Rev.\ D {\bf 93} (2016) no.6,  069901]  
[arXiv:1507.08264 [hep-th]].

\bibitem{Hasegawa:2015bza}  
F.~Hasegawa and Y.~Yamada,  
JHEP {\bf 1510} (2015) 106  
[arXiv:1507.08619 [hep-th]]. 


\bibitem{Ferrara:2015gta} 
  S.~Ferrara, M.~Porrati and A.~Sagnotti,
  Phys.\ Lett.\ B {\bf 749}, 589 (2015)
  [arXiv:1508.02939 [hep-th]].

\bibitem{Antoniadis:2015ala} 
  I.~Antoniadis and C.~Markou,
  Eur.\ Phys.\ J.\ C {\bf 75}, no. 12, 582 (2015)
  [arXiv:1508.06767 [hep-th]].




\bibitem{DallAgata:2015pdd} 
  G.~Dall'Agata, S.~Ferrara and F.~Zwirner,
  Phys.\ Lett.\ B {\bf 752}, 263 (2016)
  [arXiv:1509.06345 [hep-th]].




\bibitem{Bandos:2015xnf} 
  I.~Bandos, L.~Martucci, D.~Sorokin and M.~Tonin,
  JHEP {\bf 1602}, 080 (2016)
  [arXiv:1511.03024 [hep-th]].




\bibitem{Farakos:2016hly} 
  F.~Farakos, A.~Kehagias, D.~Racco and A.~Riotto,
  JHEP {\bf 1606}, 120 (2016)
  [arXiv:1605.07631 [hep-th]].



\bibitem{Cribiori:2016qif} 
  N.~Cribiori, G.~Dall'Agata, F.~Farakos and M.~Porrati,
  Phys.\ Lett.\ B {\bf 764}, 228 (2017)
  [arXiv:1611.01490 [hep-th]].



\bibitem{Bandos:2016xyu} 
  I.~Bandos, M.~Heller, S.~M.~Kuzenko, L.~Martucci and D.~Sorokin,
  JHEP {\bf 1611}, 109 (2016)
  [arXiv:1608.05908 [hep-th]].
  
  

\bibitem{Benakli:2017whb} 
  K.~Benakli, Y.~Chen, E.~Dudas and Y.~Mambrini,
  arXiv:1701.06574 [hep-ph].


 

\bibitem{Antoniadis:2014oya} 
  I.~Antoniadis, E.~Dudas, S.~Ferrara and A.~Sagnotti,
  Phys.\ Lett.\ B {\bf 733}, 32 (2014) 
  [arXiv:1403.3269 [hep-th]].



\bibitem{Ferrara:2014kva} 
  S.~Ferrara, R.~Kallosh and A.~Linde,
  JHEP {\bf 1410}, 143 (2014)
  [arXiv:1408.4096 [hep-th]].



\bibitem{Kallosh:2014via}  
R.~Kallosh and A.~Linde,  
JCAP {\bf 1501} (2015) 025  
[arXiv:1408.5950 [hep-th]]. 


\bibitem{DallAgata:2014qsj} 
  G.~Dall'Agata and F.~Zwirner,
  JHEP {\bf 1412}, 172 (2014)
  [arXiv:1411.2605 [hep-th]].


\bibitem{Kahn:2015mla}  Y.~Kahn, D.~A.~Roberts and J.~Thaler,
  JHEP {\bf 1510} (2015) 001
  [arXiv:1504.05958 [hep-th]].

\bibitem{Ferrara:2015tyn} 
  S.~Ferrara, R.~Kallosh and J.~Thaler,
  Phys.\ Rev.\ D {\bf 93}, no. 4, 043516 (2016)
  [arXiv:1512.00545 [hep-th]].


\bibitem{Carrasco:2015iij} 
  J.~J.~M.~Carrasco, R.~Kallosh and A.~Linde,
  Phys.\ Rev.\ D {\bf 93}, no. 6, 061301 (2016)
  [arXiv:1512.00546 [hep-th]].





\bibitem{DallAgata:2015zxp} 
  G.~Dall'Agata and F.~Farakos,
  JHEP {\bf 1602}, 101 (2016)
  [arXiv:1512.02158 [hep-th]].



\bibitem{Dudas:2016eej} 
  E.~Dudas, L.~Heurtier, C.~Wieck and M.~W.~Winkler,
  Phys.\ Lett.\ B {\bf 759}, 121 (2016)
  [arXiv:1601.03397 [hep-th]].



\bibitem{Kallosh:2016ndd} 
  R.~Kallosh, A.~Linde and T.~Wrase,
  JHEP {\bf 1604}, 027 (2016) 
  [arXiv:1602.07818 [hep-th]].


\bibitem{McDonough:2016der} 
  E.~McDonough and M.~Scalisi,
  JCAP {\bf 1611}, no. 11, 028 (2016)
  [arXiv:1609.00364 [hep-th]].

\bibitem{Hasegawa:2017hgd} 
  F.~Hasegawa, K.~Mukaida, K.~Nakayama, T.~Terada and Y.~Yamada,
  Phys.\ Lett.\ B {\bf 767}, 392 (2017)
  [arXiv:1701.03106 [hep-ph]].








	
\bibitem{Antoniadis:1999xk}  I.~Antoniadis, E.~Dudas and A.~Sagnotti,
  Phys.\ Lett.\ B {\bf 464} (1999) 38
  [hep-th/9908023].
  
\bibitem{Angelantonj:1999jh}  C.~Angelantonj,
  Nucl.\ Phys.\ B {\bf 566} (2000) 126
  [hep-th/9908064].

\bibitem{Aldazabal:1999jr}  G.~Aldazabal and A.~M.~Uranga,
  JHEP {\bf 9910} (1999) 024
  [hep-th/9908072].
  
\bibitem{Dudas:2000nv}  E.~Dudas and J.~Mourad,
  Phys.\ Lett.\ B {\bf 514} (2001) 173
  [hep-th/0012071].
  
\bibitem{Pradisi:2001yv}  G.~Pradisi and F.~Riccioni,
  Nucl.\ Phys.\ B {\bf 615} (2001) 33
  [hep-th/0107090].









\bibitem{Bergshoeff:2015jxa} 
  E.~A.~Bergshoeff, K.~Dasgupta, R.~Kallosh, A.~Van Proeyen and T.~Wrase,
  JHEP {\bf 1505}, 058 (2015)
  [arXiv:1502.07627 [hep-th]].

\bibitem{Kallosh:2015nia} 
  R.~Kallosh, F.~Quevedo and A.~M.~Uranga,
  JHEP {\bf 1512}, 039 (2015)
  [arXiv:1507.07556 [hep-th]].


\bibitem{Garcia-Etxebarria:2015lif}  I.~Garcia-Etxebarria, F.~Quevedo and R.~Valandro,
  JHEP {\bf 1602} (2016) 148
  [arXiv:1512.06926 [hep-th]].
  


\bibitem{Dasgupta:2016prs}  K.~Dasgupta, M.~Emelin and E.~McDonough,
  Phys.\ Rev.\ D {\bf 95} (2017) no.2,  026003
  [arXiv:1601.03409 [hep-th]].




\bibitem{Vercnocke:2016fbt} 
  B.~Vercnocke and T.~Wrase,
  JHEP {\bf 1608}, 132 (2016)
  [arXiv:1605.03961 [hep-th]].

\bibitem{Kallosh:2016aep} 
  R.~Kallosh, B.~Vercnocke and T.~Wrase,
  JHEP {\bf 1609}, 063 (2016)
  [arXiv:1606.09245 [hep-th]].


\bibitem{Aalsma:2017ulu} 
  L.~Aalsma, J.~P.~van der Schaar and B.~Vercnocke,
  arXiv:1703.05771 [hep-th].









 
 
 
 
 
\bibitem{Volkov:1973ix}  D.~V.~Volkov and V.~P.~Akulov,  
Phys.\ Lett.\  {\bf 46B} (1973) 109.
  



\bibitem{IK1} 
E. A. Ivanov and A. A. Kapustnikov, 
J. Phys. A \textbf{11} (1978) 2375. 




\bibitem{Samuel:1982uh}   
S.~Samuel and J.~Wess,  
Nucl.\ Phys.\ B {\bf 221}, 153 (1983). 
 
 

\bibitem{Cribiori:2016hdz}
  N.~Cribiori, G.~Dall'Agata and F.~Farakos,
  Phys.\ Rev.\ D {\bf 94} (2016) no.6,  065019
  [arXiv:1607.01277 [hep-th]].
  


\bibitem{Kuzenko:2011tj} 
  S.~M.~Kuzenko and S.~J.~Tyler,
  JHEP {\bf 1105}, 055 (2011)
  [arXiv:1102.3043 [hep-th]].







\bibitem{Kuzenko:2011ti} 
  S.~M.~Kuzenko and S.~J.~Tyler,
  JHEP {\bf 1104}, 057 (2011)
  [arXiv:1102.3042 [hep-th]].

\bibitem{Farakos:2013zsa} 
  F.~Farakos, S.~Ferrara, A.~Kehagias and M.~Porrati,
  Nucl.\ Phys.\ B {\bf 879}, 348 (2014)
  [arXiv:1309.1476 [hep-th]].

\bibitem{Farakos:2015vba} 
  F.~Farakos, O.~Hulik, P.~Koci and R.~von Unge,
  JHEP {\bf 1509}, 177 (2015)
  [arXiv:1507.01885 [hep-th]].

\bibitem{Koci:2016rqf} 
  P.~Koci, K.~Koutrolikos and R.~von Unge,
  JHEP {\bf 1702}, 076 (2017)
  [arXiv:1612.08706 [hep-th]].

\bibitem{Delacretaz:2016nhw} 
  L.~V.~Delacretaz, V.~Gorbenko and L.~Senatore,
  JHEP {\bf 1703}, 063 (2017)
  [arXiv:1610.04227 [hep-th]].



\bibitem{Ferrara:2016ajl} 
  S.~Ferrara, A.~Kehagias and A.~Sagnotti,
  Int.\ J.\ Mod.\ Phys.\ A {\bf 31}, no. 25, 1630044 (2016)
  [arXiv:1605.04791 [hep-th]].



 




\end{thebibliography}


}
\end{document}